# Little Boxes: The Simplest Demonstration of the Failure of Einstein's Attempt to Show the Incompleteness of Quantum Theory

John D. Norton[a]

The failure of Einstein's co-authored "EPR" attempt to show the incompleteness of quantum theory is demonstrated directly for spatial degrees of freedom using only elementary notions. A GHZ construction is realized in the position properties of three particles whose quantum waves are distributed over three two-chambered boxes. The same system is modeled more realistically using three spatially separated, singly ionized hydrogen molecules.

## I. INTRODUCTION

In a celebrated critique, Einstein,[1] as co-author and single author, used entangled systems in an attempt to show that the quantum state gives an incomplete description of the properties really possessed by a quantum system. There are, he urged, further hidden properties. In a celebrated rejoinder, Bell,[2] and those who developed his approach, showed that no theory employing hidden states that are spatially separate and local could recover the same predictions as quantum theory.

---

[a] Department of History and Philosophy of Science and Center for Philosophy of Science, University of Pittsburgh, Pittsburgh PA 15260. http://www.pitt.edu/~jdnorton

[1] A. Einstein, B. Podolsky, N. Rosen, "Can Quantum-Mechanical Description of Physical Reality Be Considered Complete?" Phys. Rev. **47**, 777–780 (1935) [henceforth "EPR"]; A. Einstein, "Physics and Reality," Journal of the Franklin Institute **221**, 349-382 (1936); A. Einstein, "Autobiographical Notes" in *Albert Einstein: Philosopher-Scientist*, edited by P. A. Schilpp (1949, 2nd ed. New York, Tudor, 1951), pp. 1-94; A. Einstein, "Quanten-Mechanik und Wirklichkeit," Dialectica, **2**, 320-24 (1948).

[2] J. S. Bell, "On the Einstein-Podolsky-Rosen Paradox", *Physics*, **1** 195-200 (1964).

The literature has followed Bell in developing its analyses in terms of the internal degrees of freedom associated with spin, for they are mathematically the simplest and lend themselves to experimental test.[3] These analyses preclude further, hidden spin properties. What of the familiar, spatial degrees of freedom discovered early in the development of quantum theory that are associated with the spread of the quantum wave?

This paper will show how these same analyses can be applied directly to spatial degrees of freedom and that, analogously, they preclude further hidden, position properties. In so far as is possible, the demonstration will employ only the most elementary notions: that a quantum system can be represented by a wave and that measurement collapses the wave. More specifically, the analysis will adapt Bernstein's[4] simplified analysis of a GHZ construction to spatial degrees of freedom.

The spatial degrees of freedom of quantum systems will be introduced through an example that Einstein communicated to Schrödinger shortly after the EPR paper had been written. While trying to explain the essential ideas of the EPR paper, Einstein illustrated the notion of incompleteness with the simple idea of a ball that will assuredly appear in one or other of two boxes. Since the ball's quantum wave is spread over the two boxes, Einstein urged that the wave provides an incomplete description of the ball. Einstein's account will be given in Section II below and a version of his system, as a two-chambered box with a quantum particle, will be described in Section III. A realistic implementation of the system is recovered from molecular orbital chemistry through the singly ionized hydrogen molecule, $H_2^+$.

---

[3] While none of Einstein's published versions of his attempt to show incompleteness employed spin degrees of freedom, Tilman Sauer has recently found an unpublished Einstein manuscript of late 1954 or early 1955 in which the analysis is developed in terms of spin. T. Sauer, "An Einstein Manuscript on the EPR Paradox for Spin Observables," Studies in History and Philosophy of Modern Physics, **38**, 879-887 (2007).

[4] H. J. Bernstein, "Simple Version of the Greenberger-Horne-Zeilinger (GHZ) Argument Against Local Realism," Foundations of Physics, **29**, 521-25 (1999).

Einstein's attempt to show incompleteness depended essentially on the properties of entangled systems. These properties will be described briefly in Section IV. Einstein's coauthored "EPR" analysis, sketched in Section V, uses measurements of the properties of one quantum system in an attempt to discern the properties really possessed by another spatially distant quantum system entangled with it. Section VI describes a system of three entangled particles, spread over three two-chambered boxes. The hidden position property sought by Einstein's methods is whether each of the boxes' particles is in the left or the right chamber.

In Section VII, we shall see how possible measurements on any two of the boxes enables remote determination of the position properties of the third. The totality of such measurements constrains the position properties that must be possessed by the quantum particles. Two sorts of measurements will be described. One is just a straightforward position measurement concerning which chamber hosts the particle (Sections VII.1, VII.2). The other is a determination of whether the particle's wave is spread over the two chambers in a "bonding" or "antibonding" state, drawing here on a notion from molecular orbital theory (Sections VII.3, VII.4). It turns out that the two sets of measurements lead to contradictory ascriptions of the hidden position properties. That establishes the failure of Einstein's attempt to show the incompleteness of quantum theory.

## II. EINSTEIN'S BOXES[5]

The EPR paper was received by *Physical Review* on March 25, 1935, and published on the following May 15. Shortly after, in a letter of June 19, 1935,[6] Einstein expressed his dissatisfaction with the exposition. "For reasons of language," Einstein explained to Schrödinger, "[it] was written by Podolosky after many discussions. But still it has not come out as well as I really wanted; on the contrary, the main point was, so to speak, buried by the erudition."[7] He then identified the real problem as residing in the meaning of the assertions that a description of reality is complete or incomplete. To explain this, Einstein introduced the illustration from which this note will depart:

> …I want to explain the[se assertions] through the following illustration:
>> In front of me are two boxes with hinged lids, into which I can see when the lids are opened; the latter is called "making an observation." There is also a ball there that will always be found in one or other of the boxes, if one makes an observation.

---

[5] T. Norsen, "Einstein's Boxes," American Journal of Physics, **73**, 164-176 (2005), also develops the example of Einstein's boxes. However, he considers only one quantity, position, as opposed to the non-commuting quantities of the present analysis; and arrives at weaker results, as pointed out by A. Shimony, "Comment on Norsen's Defense of Einstein's 'Box Argument'," American Journal of Physics, **73**, 177-178 (2005).

[6] Einstein Archive, Document 22-047. Portions of this letter have been reproduced in German and in English translation in work by Arthur Fine and Don Howard. My translation is a variation of them. A. Fine, *The Shaky Game: Einstein, Realism and the Quantum Theory*. (Chicago: University of Chicago Press 1986), p. 69; D. Howard, "Einstein on Locality and Separability," *Studies in History and Philosophy of Science*, **16** 171-201 (1985).

Einstein then continued to distinguish his view of the incompleteness of quantum description ("no," below) from the standard view of completeness ("yes"):

> I now describe a state as follows: *The probability that the ball is in the first box is 1/2.* - Is this a complete description?
>
> *No*: A complete statement is: the ball *is* in the first box (or is not). That, therefore, is how the characterization of the state must appear in a complete description.
>
> *Yes*. Before I open the box, the ball is by no means in *one* of the two boxes. Being in a definite box only comes about when I lift the lid….
>
> (Einstein's emphasis)

## III. A PARTICLE CONFINED TO A TWO-CHAMBERED BOX

The following will take Einstein's example of a ball that is quantum mechanically spread over two boxes as its basic unit. To keep the analysis tractable, Einstein's two boxes will be joined together as two spatially separated compartments or chambers "L" (left) and "R" (right) of one larger box, as shown in Figure 1; and Einstein's ball will be a single quantum particle.

---

[7] Translation from Howard, Ref. 6.

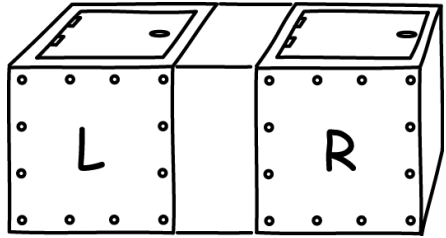

Figure 1. A Two-Chambered Box

The simplest quantum mechanical implementation of this system is a single particle confined by two infinitely deep square potential wells.[8] Figure 2 shows one of many possible wave functions.

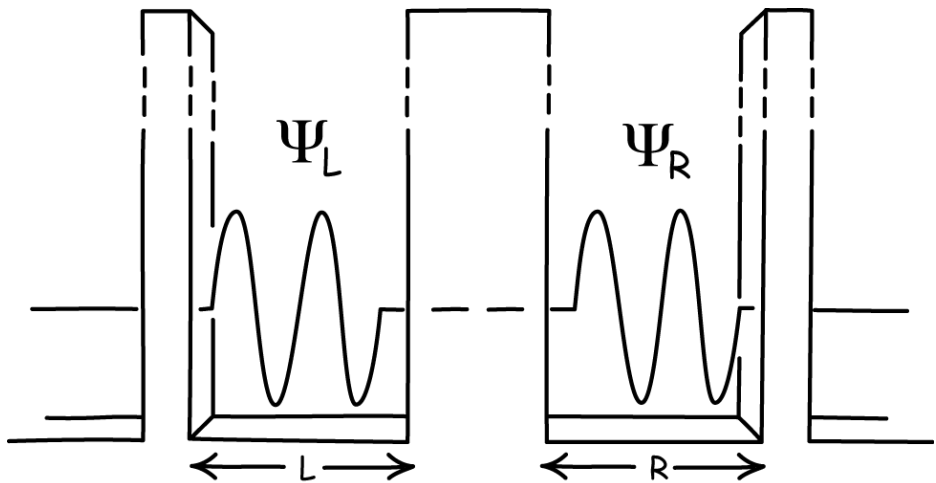

Figure 2. Particle in Two Infinitely Deep Square Potential Wells.

The two quantum waves $\Psi_L$ and $\Psi_R$ are normalized solutions of the Schrödinger equation for the case of each potential well individually. The combined wave is their normalized sum

$$\Psi_\beta = (1/\sqrt{2})(\Psi_L + \beta\Psi_R). \tag{1}$$

where $\beta$ is a relative internal phase factor of unit norm, such as 1, -1, i, -i, etc.

    A more realistic implementation of the system is based on a singly ionized hydrogen molecule, $H_2^+$. This molecule consists of two protons, whose positive charges create a potential well with two chambers centered on each proton. The single electron of the ionized molecule is

---

[8] For extended treatment of the solution of Schrödinger's equation for square potentials, see D. Bohm, *Quantum Theory*. (Englewood Cliffs, New Jersey: Prentice Hall, 1951; reprinted Mineola, N. Y.: Dover, 1989), Ch. 11.

spread across these two chambers. In the standard approximation of molecular orbital theory, LCAO "linear combination of atomic orbitals"[9], the molecule is represented, as shown in Figure 3, as consisting of a linear superposition of two 1s electron orbitals of the ordinary hydrogen atom.

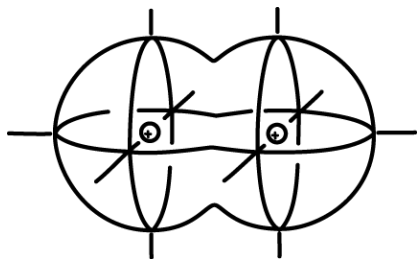

Figure 3. Singly ionized hydrogen molecule $H_2^+$

The σ bond that joins the two atoms results from the overlapping of these two orbitals. It will be important for the analysis that follows that the compartments be disjoint spatially. So the relevant system consists of an $H_2^+$ molecule in which the two nuclei have been separated sufficiently in space so that the two 1s orbitals no longer overlap, as shown in Figure 4.

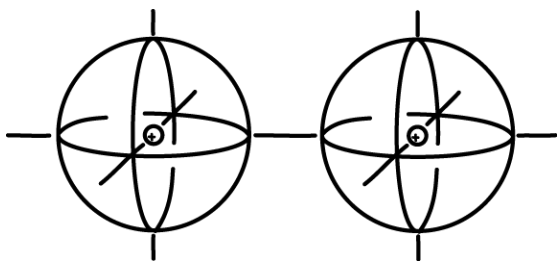

Figure 4. Separated singly ionized $H_2^+$ molecule

It turns out that overlapping is avoided almost completely by merely separating the two nuclei by about ten Bohr radii[10]. If the two 1s orbitals of the separated $H_2^+$ molecule are written as $\Psi_L$ and $\Psi_R$, then the wave of the molecule's single electron is given by Eq. (1) above.

---

[9] C. J. Ballhausen and H. B. Gray, *Molecular Orbital Theory*. (New York: W. A. Benjamin, 1965), Ch. 2.1.

[10] P. Atkins, *Physical Chemistry*. (New York: W. H. Freeman, 1994), p. 479, Fig. 14.18.

Virtually all the details of these last two instantiations of the particle in a two-chambered box will be immaterial to the analysis that follows. All that matters is that we have a quantum particle whose wave function is non-vanishing only in two disjoint regions of space that we will label "L" and "R." The wave functions of the particle, were it confined to each region individually, are $\Psi_L$ and $\Psi_R$. In general, many possible quantum waves are admissible for $\Psi_L$ and $\Psi_R$. In the case of a square potential well, they correspond to the different energy eigenstates of a particle trapped in such a potential; and, in the case of separated $H_2^+$, the various orbitals of a hydrogen atom. Which one is chosen will be immaterial. All that matters is that we fix on a particular pair of wave functions $\Psi_L$ and $\Psi_R$ and use them alone for the remainder of the analysis. As before, the wave function for the particle spread over the two boxes is given by Eq. (1) above.

## IV. ENTANGLEMENT AND THE MEASUREMENT OF DISTANT PARTICLE STATES

The EPR analysis depends essentially on the entanglement of systems. In the case of spatial degrees of freedom, the properties of entanglement in quantum measurement are readily recovered from the notion of the spatial collapse of the wavepacket under quantum measurement.

The simplest case arises when we have two two-chambered boxes, A and B, shown in Figure 5. We will assume that the two boxes are very widely separated in space. One may be on earth; the other may be light years away, resting comfortably on another planet, orbiting a distant star.

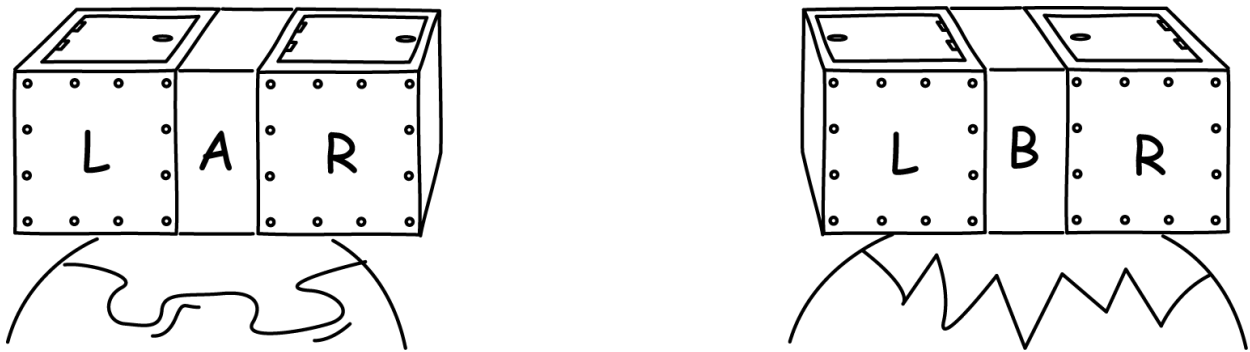

Figure 5. Two Two-Chambered Boxes

Nonetheless, the particles of each of these boxes may be entangled and in different ways. One simple way is that the A particle's presence in its L chamber is associated with the B particle's presence in its L chamber; and similarly for the R chambers. The wave of the resulting entangled state is written as

$$\Psi_{AB} = (1/2)(\Psi_{A,L}\Psi_{B,L} + \Psi_{A,R}\Psi_{B,R}). \tag{2a}$$

Each of the waves $\Psi_{A,L}, \Psi_{B,L}, \Psi_{A,R}$ and $\Psi_{B,R}$ are waves in a three dimensional space. The waves $\Psi_{A,L}$ and $\Psi_{A,R}$ are waves in the position space of particle A. The waves $\Psi_{B,L}$ and $\Psi_{B,R}$ are waves in the position space of particle B. The essential point about the wave $\Psi_{AB}$ is that it is not a wave in a three dimensional space. It is a wave in the *six* dimensional configurations space that is formed as the product of the two spaces of particles A and B.

The wave $\Psi_{AB}$ is shown in Figure 6 in its configuration space, where one dimension of space for each of the A and B particles is shown and the remaining four dimensions are suppressed.

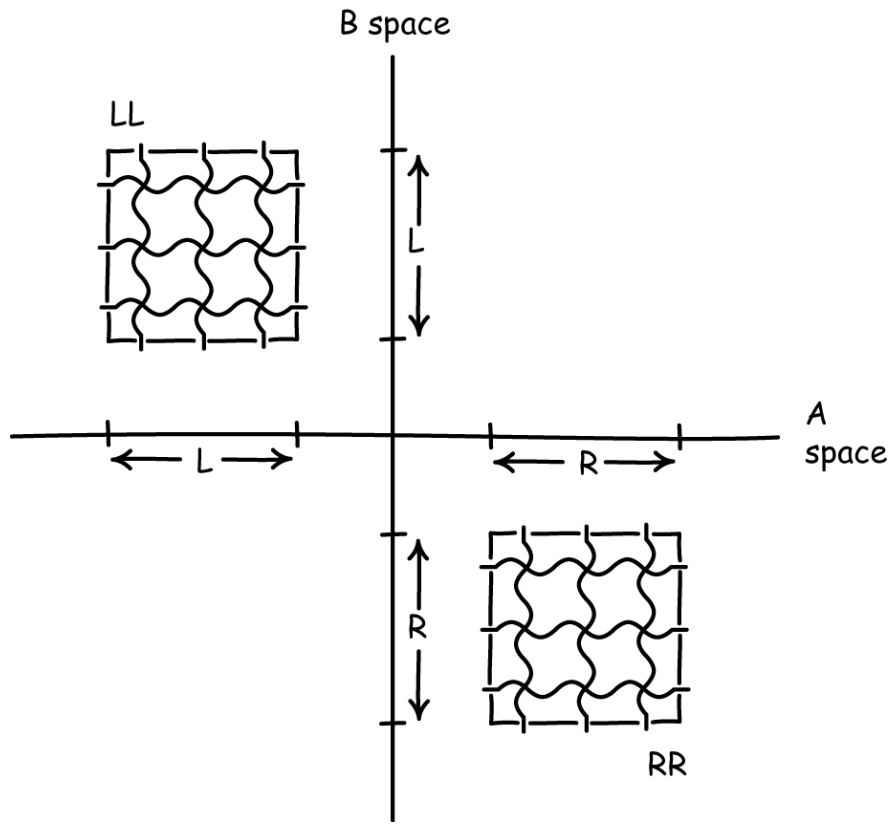

Figure 6. Configuration Space of Two Correlated Particles of Boxes A and B

The wave is non-vanishing only in two regions of the configuration space. One coincides with the L regions of both A and B spaces, "LL", and corresponds to the product term $\Psi_{A,L}\Psi_{B,L}$ of (2a). The other coincides with the R regions of both A and B spaces, "RR," and corresponds to the product term $\Psi_{A,R}\Psi_{B,R}$ of Eq. (2a).

Assume that we perform a measurement on the A particle that determines whether the particle is in the L or the R chamber. Under the standard interpretation of quantum theory, the effect of that measurement is to collapse the wave to one that is non-vanishing only in one of the L or R regions of A's space. That is, the measurement will collapse the wave $\Psi_{AB}$ to one or other of the regions LL or RR of the configuration space. If the outcome of the measurement on the A particle was L, then the wave would have collapsed to the LL region. This collapse would alter the B wave, which would now only be non-vanishing in its L region. That is, the wave of Eq. (2a) has collapsed to just one of its terms

$$\Psi_{AB} = \Psi_{A,L}\Psi_{B,L}. \tag{2b}$$

As a result, we can now be certain that any subsequent measurement of the position of the B particle will reveal it to be in its L region.

This is the important fact about entanglement for present purposes. Even though we have measured the A particle only, we now are certain about the outcomes of measurements were they to be performed on the B particle, light years away.

## V. THE EPR ANALYSIS

Einstein's repeated objection to quantum theory was that it was incomplete under its standard interpretation. That is, specifying the quantum wave associated with a particle does not, he believed, fix all the physical properties of the particle. If the particle's wave is spread over the two chambers of the box, then standard quantum theory tells us that the particle is neither properly in one or other of the chambers. Einstein, however, urged that the particle has further factual properties, not captured in the quantum wave, that correspond to its localization in one or other of the chambers.

Einstein's well-known strategy for establishing this incompleteness was first advanced in the coauthored "EPR" paper and then under his name only in some of Einstein's later publications.[11] The analysis depended essentially on the behavior of entangled systems under measurement, as sketched in the last section. Here is how it is implemented in that example. Measurement of the position of the A particle enables us to know what a position measurement of the B particle would assuredly yield, even though the B particle is so remote from our measurement that, presumably, our measurement could have no effect on it. This fact, it is asserted, establishes that the B particle had a definite position property all along, even though the original quantum wave did not express it.

The analysis depends upon an inference from our being able to predict assuredly the outcome of a measurement on the B particle, to the possession by the B particle of a property. The assumption that justified that inference was of central importance to the EPR analysis and thus stated in italics on the first page of the EPR paper:

> [EPR reality criterion]
>
> If, without in any way disturbing a system, we can predict with certainty (i.e. with probability equal to unity) the value of a physical quantity, then there exists an element of physical reality corresponding to this physical quantity.

There is an important counterfactual element in the EPR analysis.[12] It does not actually require that we perform the measurements on the A particle to infer to the possession of position properties by the B particle. All that matters is that we know what *would* happen if, counter to the facts, we were to perform the measurement. If we could predict assuredly that there would be some definite position measured on the remote B particle, then we can infer that the remote particle possesses some definite position property. All we lose by not performing a measurement on the A particle is that we do not know which position property is possessed by the remote B particle. Since we need not actually perform the measurements, analogous reflections on the possibility of measurements on the B particle lead to the corresponding conclusion for the A particle: it too possesses a definite position property not expressed completely by its quantum wave. The counterfactual aspect is essential if we are to infer that *both* the A and B particle possess more position properties than expressed by their quantum waves. For, we can perform the measurement on at most one of the two entangled particles, since, under the standard theory, measurement on one alters the combined system.[13]

---

[11] Ref. 1.

[12] This allowance of counterfactual measurement appears in the EPR paper in the consideration of measurements on the same system of observables corresponding to non-commuting operators, such as, in their example, the position and momentum of one of the particles. Both measurements are considered, but at most one could be carried out.

The literature has subjected the presumptions of the EPR analysis to a dissection more minute than can be recapitulated here. Howard[14] has traced the presence of two important presumptions in Einstein's thought. First is separability: the assumption that the two spatially remote systems have independent existence. Second is locality: the assumption that effects between the systems propagate at luminal speeds or less. Briefly, the first affirms that the remote particle has properties independent of the local particle measured; and the second assures us that whatever disturbances are introduced by measurement will initially affect only the local particle.

## VI. THREE TWO-CHAMBERED BOXES, ENTANGLED

The failure of the EPR attempt to show the incompleteness of quantum theory was demonstrated by Bell[15] and work that developed his approach in many different ways. The failure is demonstrated by showing that no physical theory, constrained by separability and locality, can reproduce the empirical predictions of quantum theory. Most analyses follow Bell in considering two entangled systems, such as two spin half particles in a singlet state. In that case, the empirical failure of separable, local theories emerges in their failure to reproduce the correlations between measurement outcomes predicted by quantum theory for repeated trials.

---

[13] Bohr's celebrated response to the EPR paper depends on denying this counterfactual and insisting that, were different measurements performed, the system would somehow be different. As a result, considering different measurements counterfactually on the first system no longer enables us to extend the list of properties we infer as possessed by the remote system. (For a mature version, see N. Bohr, "Discussions with Einstein on Epistemological Problems in Atomic Physics," in *Albert Einstein-Philosopher Scientist*. 2nd ed., edited by P. A. Schilpp (New York: Tudor Publishing, 1951), pp. 201-241.) Assessments of Bohr's response to the EPR paper vary widely. In my view, Bohr's response fails since it depends upon an extreme form of empiricism that conflates what a system is with how we find out about it. Standard quantum theory does not make that conflation. It provides clear statements as to what would happen were, counterfactually, different measurements to be performed on one and the same quantum system.

[14] Ref. 6.

[15] Ref. 2.

A significant conceptual simplification was achieved by Greenberger, Horne and Zeilinger ("GHZ").[16] If one considers just two entangled systems, the proof of the failure of the EPR analysis works indirectly by showing that a local, separable theory cannot return the statistical correlations among measurements predicted by quantum theory. In the GHZ approach, more than two entangled systems are considered. That enables the EPR reality criterion to be applied directly. From it and the predictions of quantum theory, one can infer to the properties a spatially remote system must possess and note that the ensuing requirements are self-contradictory.

The entangled system to be developed here is a version of a GHZ system. It is the one described in Mermin[17] and simplified by Bernstein.[18] The analysis has been modified by substituting spatial degrees of freedom for the spin degrees of freedom.[19] The system consists of

---

[16] D. Greenberger, M. Horne, A. Zeilinger, "Going Beyond Bell's Theorem," in *Bell's Theorem, Quantum Theory, and Conceptions of the Universe*, edited by M. Kafatos (Dordrecht: Kluwer, 1989) pp. 73-76. See also R. Clifton, M. L. G. Redhead, J. Butterfield, "Generalization of the Greenberger-Horne-Zeilinger Algebraic Proof of Nonlocality," Foundations of Physics, **21** 149-84 (1991).

[17] D. Mermin, "What's Wrong with These Elements of Reality?" Phys. Today **43**(6) 9-11 (1990).

[18] Ref. 4.

[19] The substitution depends upon the fact that the two states $\Psi_L$ and $\Psi_R$ are orthogonal, since they are non-vanishing only in spatially disjoint regions. Hence they span a two-dimensional Hilbert space, which can be substituted for the traditional spin-1/2 spaces employed in GHZ constructions. One mapping associates states $\Psi_L$ and $\Psi_R$ with z-spin up and z-spin down; states $\Psi_i$ and $\Psi_{-i}$ of (1) with y-spin up and y-spin down; and states $\Psi_1$ and $\Psi_{-1}$ of (1) with x-spin up and x-spin down. This mapping enables a translation between a GHZ construction in terms of spin properties and the construction used in this paper.

three, spatially remote, two-chambered boxes, A, B and C, with one particle in each box, as shown in the Figure 7:

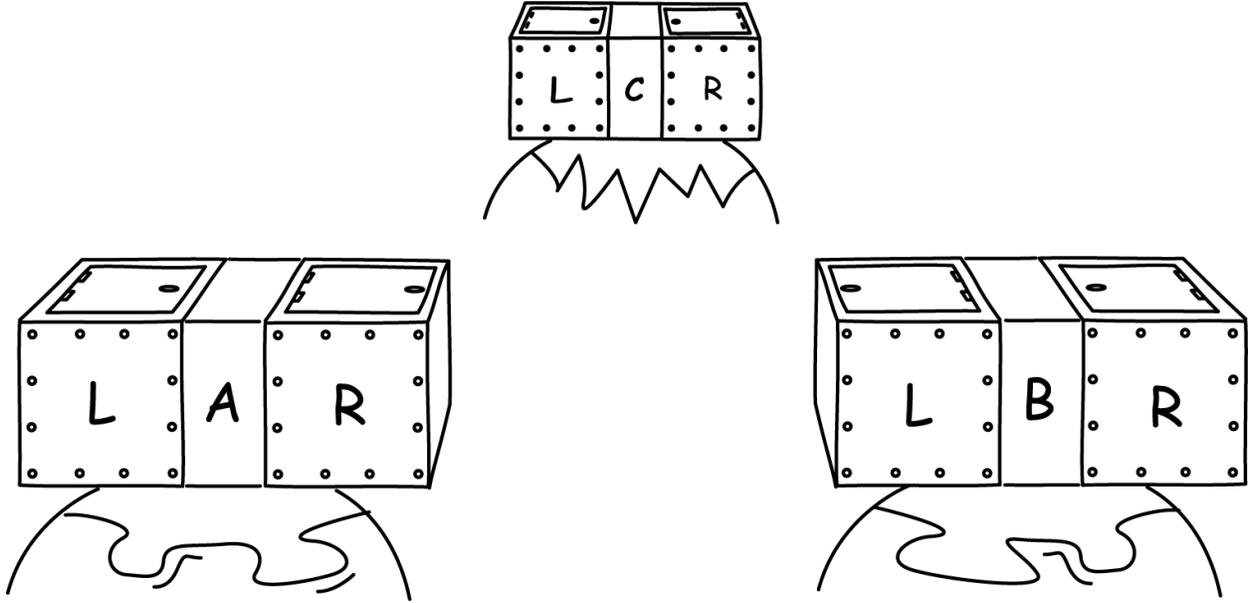

Figure 7. Three Two-Chambered Boxes

To display the failure of the EPR analysis, the entanglement of the particles in the three boxes will be a little more complicated than that the two particle system of Eq. (2a). Each particle of the boxes of A, B and C are spread over the two chambers of the box. There are two ways the A particle will be spread:

$$\Psi_{A,+i} = (1/\sqrt{2})(\Psi_{A,L} + i\Psi_{A,R}) \qquad \Psi_{A,-i} = (1/\sqrt{2})(\Psi_{A,L} - i\Psi_{A,R}). \qquad (1a)$$

The first has a relative internal phase factor $\beta$ in (1) of $+i$; and the second has a relative internal phase factor of $-i$. Waves for the B and C boxes are defined analogously. The three particles are entangled in the state

$$\Psi_{ABC} = (1/\sqrt{2}) \, (\Psi_{A,+i}\Psi_{B,+i}\Psi_{C,+i} - \Psi_{A,-i}\Psi_{B,-i}\Psi_{C,-i}). \qquad (3)$$

In the remainder of the paper, we will apply the EPR reality criterion to this state in order to determine the properties that criterion attributes to it, its "elements of physical reality," and then show that the resulting requirements are self-contradictory.

## VII. MEASUREMENT OF POSSESSED PROPERTIES

We shall ascertain the position properties of the three particles of the entangled state $\Psi_{ABC}$ by considering the outcomes of two sorts of measurements, position measurements and "bonding/antibonding" measurements. These measurements on two of the particles will enable us to predict the outcome of a position measurement on the third, remote particle. Using the EPR reality criterion, we shall thereby learn a possessed property of that remote particle. The totality of all possible measurements will place strong restrictions on the position properties that may be possessed by the three particles. It will turn out that the two different sorts of measurement lead to contradictory restrictions.

### *VII.1 First procedure: Position measurement*

The form in which $\Psi_{ABC}$ is written in Eq. (3) is not conducive to reading off these predictions, however. We arrive at an equivalent expression for $\Psi_{ABC}$ if we substitute for each term in Eq. (3) by means of the expressions for $\Psi_{A,+i}, \Psi_{A,-i}, \Psi_{B,+i}, \ldots$ in Eq. (1a). Multiplying out the terms leads to an unwieldy expression with 16 terms, most of which cancel, leaving the final expression

$$\Psi_{ABC} = (i/2)\,(\Psi_{A,L}\Psi_{B,L}\Psi_{C,R} + \Psi_{A,L}\Psi_{B,R}\Psi_{C,L} + \Psi_{A,R}\Psi_{B,L}\Psi_{C,L} - \Psi_{A,R}\Psi_{B,R}\Psi_{C,R}). \quad (3a)$$

We can read off this expression how position measurements on two particles enable us to predict the outcome of a position measurement on the other particle. For example, if position measurements of the particles of A and B both yield L, then the state collapses to the first term

$$\Psi_{ABC} = \Psi_{A,L}\Psi_{B,L}\Psi_{C,R}$$

and we know that the particle of C must return the position R on measurement. These four terms correspond to four regions in the configuration space of the particles of A, B and C, shown in Figure 8. The measurement described would collapse the wave to the cube on the top left at the rear, marked "LLR."

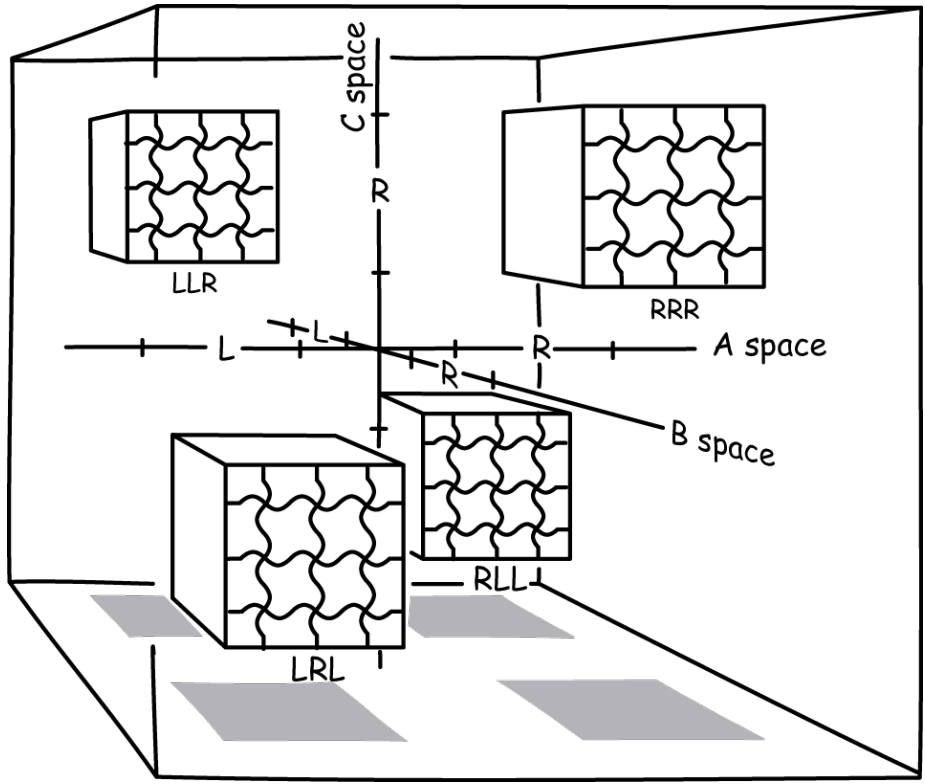

Figure 8. Configuration space of the three entangled particles of $\Psi_{ABC}$

A measurement of the positions of any two particles will collapse the wave to one of these four regions. They are

"LLR," "LRL," "RLL" and "RRR,"

where "LLR" indicates A position L, B position L and C position R, etc. We read the following rule from these four possibilities:

*Rule: "Same-R, Different-L"*

If we measure two particles to have the *same* position,

then the remaining particle will return position R on measurement.

If we measure two particles to have *different* positions,

then the remaining particle will return position L on measurement.

where "same" just means both L or both R; and "different" means one L and one R.

### *VII.2 Applying the EPR reality criterion*

We now apply the EPR reality criterion to these results that, so far, only concern measurement. When the rule predicts assuredly what position property would be measured for a remote particle, we infer that the particle possesses it as a hidden position property.

We could choose to measure the particle positions of any pair of systems: the pair BC; or the pair AC; or the pair AB. Then, the above rule[20] would enable to us predict with certainty what we would measure for the position of the particle of the third system: the particle of A, of B and of C, respectively. Recall that the EPR analysis allows us to assemble conclusions that derive from considering many different measurements that, counterfactually, are not performed. Hence we can pool all these outcomes and conclude that each of the particles of A, B and C really possess the property of L/R position.

If we consider the possible position measurements of the three possible pairings of the particle, we may ask if the members of each pair are the same or different. There are only two cases:

(a) All *three* pairs have the *same* position.

(b) *One* pair has the *same* position and the remaining *two* pairs have *different* positions.

The position properties of these two cases must conform to the Rule: "Same-R, Different-L." Applying it to case (a) tells us that all three position properties must all be R. Applying the rule to case (b) tells us two position properties are L and one is R. That is, there are only four possible distributions of position properties

$$\text{LLR, LRL, RLL and RRR,} \qquad (4a)$$

where, as before, LLR designates the properties of the A, B and C particles in that order, etc.

### *VII.3 Second procedure: Bonding/antibonding measurement*

A position measurement determines whether a particle in the two-chambered box is in the left or right chamber, that is, whether the particle manifests as $\Psi_L$ or $\Psi_R$ on measurement. A second measurement determines whether the particle manifests as $\Psi_{+1}$ or $\Psi_{-1}$, that is, as the normalized sum or difference of $\Psi_L$ and $\Psi_R$:

$$\Psi_{+1} = (1/\sqrt{2})(\Psi_L + \Psi_R) \quad \Psi_{-1} = (1/\sqrt{2})(\Psi_L - \Psi_R). \tag{1b}$$

---

[20] The same rule applies to all cases because of the manifest symmetry of the state of Eq. (3) in the three particles of A, B and C.

The waves corresponding to the two cases are shown in Figure 9 for the case of a particle spread over two square wave potential well. $\Psi_{+1}$ is on the left and $\Psi_{-1}$ on the right.

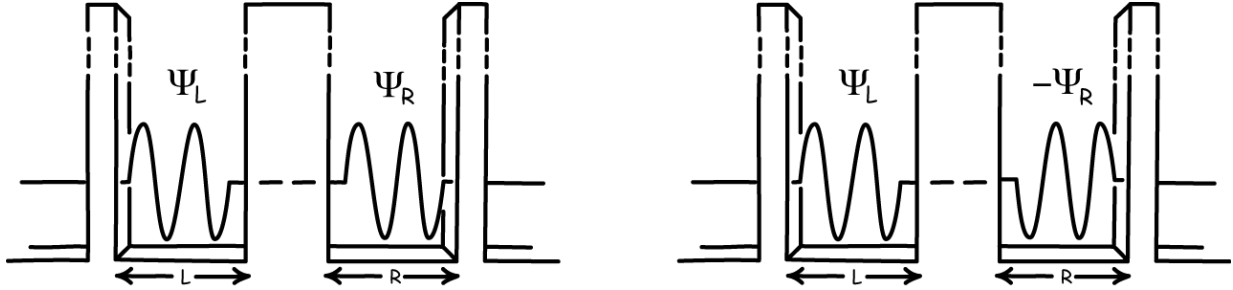

Figure 9. $\Psi_{+1}$ and $\Psi_{-1}$ for Two Square Wave Potential Wells.

From a formal perspective, it is appropriate to consider this a measurement, for the two states, $\Psi_{+1}$ or $\Psi_{-1}$, are eigenstates of a self-adjoint operator.[21] Merely satisfying this formal property may raise doubt as to whether this measurement is the sort of measurement one might actually perform physically. The states $\Psi_{+1}$ or $\Psi_{-1}$ differ only in a relative, internal phase factor. Is this a difference that measurement can reveal; or is it merely an unphysical gauge freedom?

We can put these doubts to rest. Recall that one instantiation of the two-chambered box is a separated, singly ionized hydrogen molecule $H_2^+$. In this context, the two states have distinct physical meanings.

For the ionized hydrogen molecule, the waves $\Psi_L$ and $\Psi_R$ correspond to 1s electron orbitals of the ordinary hydrogen atom. In the case of $\Psi_{+1}$, the two parts of the electron wave are the same 1s orbitals in phase. If the two hydrogen nuclei are allowed to approach, then molecular orbital theory, through the LCAO "linear combination of atomic orbitals" approximation, tells us that they form a stable lower energy state that comprises the σ bond. The bonding depends essentially on the zero phase difference (the "+1" factor) between the two waves $\Psi_L$ and $\Psi_R$. In the region of overlap between the two nuclei, because of this zero phase difference, the two waves reinforce increasing the wave amplitude between the nuclei.[22]

---

[21] As noted in an earlier footnote, the two orthogonal states, $\Psi_L$ and $\Psi_R$ span a two-dimensional Hilbert space. Since $\Psi_{+1}$ or $\Psi_{-1}$ are also orthogonal, so that $|\Psi_{+1}\rangle\langle\Psi_{-1}| = |\Psi_{-1}\rangle\langle\Psi_{+1}| = 0$, the requisite self adjoint operator acting on this space with eigenvectors $|\Psi_{+1}\rangle$ or $|\Psi_{-1}\rangle$ is just $|\Psi_{+1}\rangle\langle\Psi_{+1}| - |\Psi_{-1}\rangle\langle\Psi_{-1}|$.

We find the opposite behavior for the case of $\Psi_{-1}$. Then the two 1s orbitals are $180^0$ degrees out of phase (the "-1" factor). As a result, where the two waves overlap they will destructively interfere and cancel one another. The computations show that this is a higher energy state.[23] It is the opposite of bond; it is an unstable state in which the two nuclei repel and is known as the "antibonding" σ* state. The two states are shown in Figure 10, where the accumulation of wave amplitude is indicated by shading.

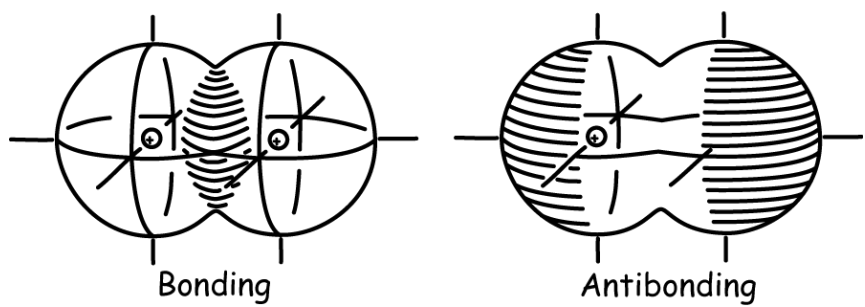

Figure 10. Bonding and Antibonding States of $H_2^+$.

---

[22] See Ref. 9 and Ref. 10, Section 14.4. For a crude heuristic picture, imagine that this means the single electron tends to inhabit the space between the nuclei more, so that its attraction for the nuclei pulls them together. Alternatively, the closer packing of the three particles (two nuclei and one electron) leads to a lower energy state.

[23] For a crude heuristic picture, this means that the three particles are more spread out and thus have a higher energy state. One might also imagine the electron more on the outside of the two nuclei, so that its attraction for the nuclei will tend to separate them.

The bonding and antibonding states enable a simple experimental procedure for determining whether the separated singly ionized hydrogen molecule $H_2^+$ is in the "bonding" state $\Psi_{+1}$ or in the "antibonding" state $\Psi_{-1}$. The nuclei are allowed to move freely. If they are in the bonding state, they will attract and approach one another; if they are in the antibonding state, they will repel.[24]

It is not easy to discern how the state $\Psi_{ABC}$ will behave under bonding/antibonding measurements, when those measurements are used to determine the position properties of a remote particle. We arrive at a more useful expression for the state if we make the following substitions. The waves $\Psi_{C,+i}$ and $\Psi_{C,-i}$ are re-expressed in terms of $\Psi_{C,L}$ and $\Psi_{C,R}$ by means of the expressions

$$\Psi_{C,+i} = (1/\sqrt{2})(\Psi_{C,L} + i\Psi_{C,R}) \qquad \Psi_{C,-i} = (1/\sqrt{2})(\Psi_{C,L} - i\Psi_{C,R}). \qquad (1a')$$

The waves $\Psi_{A,+i}, \Psi_{A,-i}, \Psi_{B,+i}$ and $\Psi_{B,-i}$ are re-expressed in terms of $\Psi_{A,+1}, \Psi_{A,-1}, \Psi_{B,+1}$ and $\Psi_{B,-1}$ by means of[25]

$$\Psi_{A,+i} = ((1+i)/2)(\Psi_{A,+1} - i\Psi_{A,-1}) \qquad \Psi_{A,-i} = ((1-i)/2)(\Psi_{A,+1} + i\Psi_{A,-1}) \qquad (1c)$$

$$\Psi_{B,+i} = ((1+i)/2)(\Psi_{B,+1} - i\Psi_{B,-1}) \qquad \Psi_{B,-i} = ((1-i)/2)(\Psi_{B,+1} + i\Psi_{B,-1}).$$

Multiplying out all the resulting terms and cancelling where possible leads to the expression

$$\Psi_{ABC} = (i/2)(\Psi_{A,+1}\Psi_{B,+1}\Psi_{C,L} - \Psi_{A,-1}\Psi_{B,-1}\Psi_{C,L} + \Psi_{A,+1}\Psi_{B,-1}\Psi_{C,R} + \Psi_{A,-1}\Psi_{B,+1}\Psi_{C,R}). \qquad (3a)$$

Proceeding as before, we can see how measurements of bonding/antibonding on the particles of A and B enable us to predict the outcome of a position measurement on the particle of C. If, for example, we measure both particles of A and B to be in the bonding state, then the wave of Eq. (3a) has collapsed to

$$\Psi_{ABC} = \Psi_{A,+1}\Psi_{B,+1}\Psi_{C,L}$$

---

[24] At the level of the LCAO approximation, the energies of the states $\Psi_{+i}, \Psi_{-i}, \Psi_L$ and $\Psi_R$ are independent of the internuclear distance, so none of these states form bonding or antibonding interactions. Hence testing for attraction and repulsion tests specifically for the states $\Psi_{+1}$ and $\Psi_{-1}$.

[25] To see that these are the correct expressions for $\Psi_{A,+i}$, etc., substitute in them for $\Psi_{A,+1}$, etc. using (1b), and thereby recover expressions (1a) for $\Psi_{A,+i}$, etc.,

and we are certain that a subsequent measurement on the particle of C will show an L position.

Since systems A, B and C enter symmetrically into the state of Eq. (3) these results will hold no matter which two are the particles on which bonding/antibonding measurements are performed. Hence, from (3a), we read off the following rule, similar to that found for position measurements

*Rule: "Same-L, Different-R"*

If we measure two particles to have the SAME bonding state,

then the remaining particle will return position L on measurement.

If we measure two particles to have DIFFERENT bonding states,

then the remaining particle will return position R on measurement.

where "same" just means both bonding or both antibonding; and "different" means one bonding and one antibonding.

The point to note is that the role of L and R in the rule for bonding measurements is reversed in comparison to the rule associated with position measurements. This reversal will yield the contradiction.

## *VII.4 Applying the EPR reality criterion*

We proceed as before by assembling the outcomes of all possible bonding measurements on pairs of particles. These bonding measurements must cohere with one another.[26] Hence there are only two cases for the various pairs of bonding measurement on the three particles:

---

[26] This assertion depends upon the further assumption that the bonding state of a particle is a property possessed by it, akin to its possession of a position property. This further assumption can be established by an EPR style analysis akin to those used to establish the reality of the position properties. Assume that we measure the bonding state of the particle of A and the position of the particle of C. Equation (3a) will then allow us to predict with certainty which bonding state the particle of B will show under measurement. If, for example the particle of A is bonding and the particle of C is L, then $\Psi_{ABC}$ collapses to the first term of Eq. (3a) and we can predict with certainty that a bonding measurement on the particle of B will show it to be bonding. Analogous analysis with other pairs leads to the conclusion that each particle possesses the property of a definite bonding state.

(a) All *three* pairs have the *same* bonding states.

(b) *One* pair has the *same* bonding state and the remaining *two* pairs have *different* bonding states.

The position properties of the remote, unmeasured particles must all conform to the Rule: "Same-L, Different R." Applying it to case (a) tells us that the three position properties must all be L. Applying the rule to case (b) tells us that one position property is L and that the other two must be R. That is, there are only four possible distributions of the position properties

$$RRL, RLR, LRR, LLL. \qquad (4b)$$

One sees immediately that these four distributions of position properties contradicts the distribution of position properties (4a) inferred from position measurements.

## VIII. CONCLUSION

We have arrived at a contradiction. Position measurements on the three entangled particles of quantum state Eq. (3) lead to the conclusion that the particles must possess position properties that contradict those inferred from bonding/antibonding measurements.

Two principal assumptions were made in the arguments that generated this contradiction. One was that the empirical predictions of quantum theory are reliable; the other was the EPR reality criterion, which in turn depended upon the assumptions of separability and locality. One of these assumptions must be given up. The continuing empirical success of quantum theory has led to a consensus that it is the second assumption, the EPR reality criterion, that is to be discarded.

## ACKNOWLEDGMENTS

I am grateful to Julia Bursten, Simon Garcia, Eugen Schwarz and Michael Weisberg for helpful guidance on chemical bonding; and I also thank Diana Buchwald, Tony Duncan, Wayne Myrvold, David Snoke and Giovanni Valente for their advice and assistance.

## List of Figures

Figure 1. A Two-Chambered Box

Figure 2. Particle in Two Infinitely Deep Square Potential Wells.